\documentclass[twocolumn,showpacs,preprintnumbers,amsmath,amssymb,superscriptaddress]{revtex4}
\usepackage{graphicx}% Include figure files
\usepackage{dcolumn}% Align table columns on decimal point
\usepackage{bm}% bold math
\usepackage{longtable}% bold math
\usepackage[countmax]{subfloat}
\usepackage{epstopdf}

\newcommand{\meanv}[1]{\left\langle#1\right\rangle}

\newcommand{\bxi}{{\bm{\xi}}}
\newcommand{\bsigma}{{\bm{\sigma}}}
\newcommand{\btau}{{\bm{\tau}}}

\newcommand{\branch}{B}

\usepackage{epsfig}
\usepackage{amssymb}
\usepackage{amsmath}
\usepackage{amsthm}
\usepackage{latexsym}
\usepackage{graphicx}
\def\be{\begin{equation}}
\def\ee{\end{equation}}
\def\bc{\begin{center}}
\def\ec{\end{center}}

\def\be{\begin{equation}}
\def\ee{\end{equation}}
\def\bea{\begin{eqnarray}}
\def\eea{\end{eqnarray}}

\begin{document}

\preprint{APS/123-QED}

\title{Extensive load in multitasking associative networks}

\author{Peter Sollich}
\affiliation{Department of Mathematics, King's College London, Strand,
  London WC2R 2LS, U.K.}
\author{Daniele Tantari}
\affiliation{Dipartimento di Matematica, Sapienza Universit\`a di Roma, P.le Aldo Moro 2, 00185, Roma, Italy.}
\author{Alessia Annibale}
\affiliation{Institute for Mathematical and Molecular Biomedicine,
  King's College London, Hodking Building, London SE1 1UL, U.K.}
\affiliation{Mathematics Department, King's College London, The Strand WC2R 2LS, London, UK.}
\author{Adriano Barra}
\affiliation{Dipartimento di Fisica, Sapienza Universit\`a di Roma, P.le A. Moro 2, 00185, Roma, Italy.}

\date{\today}

\begin{abstract}
 %Associative networks with diluted and symmetrically distributed patterns have been recently introduced and studied using replica theory, in different regimes of storage load and homogeneous (pattern-independent) dilution.
\noindent
We use belief-propagation techniques to study the
equilibrium behavior of a bipartite spin-glass, with interactions
between two sets of $N$ and $P =
\alpha N$ spins. Each spin has a finite
degree, i.e.\ number of interaction partners in the opposite set; an
equivalent view is then 
of a system of $N$ neurons storing $P$ diluted patterns.
We show that in a large part of the parameter
space of noise, dilution and storage load, delimited by a critical
surface, the network behaves as an
extensive parallel processor, retrieving all $P$ patterns {\it in
parallel} without falling into spurious states due to pattern
cross-talk and typical of the structural glassiness built into the network.
Our approach allows us to consider effects beyond those studied in
replica theory so far, including pattern asymmetry and heterogeneous
dilution. Parallel extensive retrieval is more robust for homogeneous
degree distributions, and is not disrupted by biases in the
distributions of the spin-glass links.
\end{abstract}

 \maketitle

Since their introduction in the pioneering paper by Amit Gutfreund and
Sompolinsky \cite{AGS,ags-annals}, associative neural networks (NN)
have played a central role in the statistical mechanics community,
soon becoming one of the most successful offshoots of spin glasses (SG) \cite{MPV}, as proved by the celebrated paper by Hopfield \cite{hopfield} or the excellent book by Amit
\cite{amit}.
\newline\indent
Indeed, the Hopfield model for NN \cite{hopfield} can be regarded as a
SG where the coupling between each pair of spins $\sigma_i,
\sigma_j$, $i,j=1,\ldots,N$ has the Hebbian form $J_{ij}=
N^{-1}\sum_{\mu}\xi_i^\mu\xi_j^\mu$ and the $\bxi^\mu,
\mu=1,\ldots,P=\alpha N$, represent stored patterns with entries
$\xi_i^\mu=\pm 1$ distributed as
$P(\xi_i^\mu=\pm 1)=1/2$. Replica analysis shows that,
in the thermodynamic limit, this system exhibits an ordered structure
of the free energy minima in the low noise regime and for storage load
$\alpha< \alpha_c \sim 0.14$ (within the ``replica symmetric''
ansatz). For higher loads the NN becomes equivalent to the
Sherrington-Kirkpatrick SG \cite{MPV}, as $\sqrt{N}J_{ij}$
becomes Gaussian with unit variance by a simple central limit theorem
argument \cite{guerra,Peter}.
\newline\indent
A further connection between NN and SG has been pointed out recently
in the context of bipartite SG \cite{JTB}. Consider a system of two
sets of spins, $\sigma_i, \ i=1,\ldots,N$ and $\tau_{\mu}, \
\mu=1,\ldots,P$, connected by links $\xi_i^{\mu}=\pm 1$
that are now {\it sparse} so that $P(\xi_i^{\mu}=\pm 1)=c/2N$ and
$P(\xi_i^\mu=0)=1-c/N$ with $c=\mathcal{O}(N^0)$, and described by the
SG Hamiltonian $H_{SG} (\bsigma,\btau | \bxi) \propto
-\sum_{i,\mu}\xi_i^{\mu} \sigma_i \tau_{\mu}$. Marginalizing over
$\btau$ in the partition function
%\be\small
%\hspace*{-4mm}
$Z = \sum_{\bsigma,\btau} e^{- \beta H_{SG}(\bsigma,\btau | \bxi)}
= \sum_{\bsigma}e^{- \beta H_{NN}(\bsigma | \bxi)}$
shows that the $\bsigma$ represent a NN with Hamiltonian
$
H_{NN}(\bsigma | \bxi) = -\beta^{-1}\sum_\mu \ln[2\cosh(\beta
\sum_i \xi_i^{\mu}\sigma_i)]$
or, up to an additive constant, $
H_{NN}(\bsigma | \bxi) = -\frac{\beta}{2}
\sum_{\mu,i,j}(\xi_i^{\mu}\xi_j^{\mu})\sigma_i\sigma_j + \ldots
%\label{eq:marginal}
$
Higher order interactions are not written explicitly here; these are
fully absent if the $\tau_i$ are continuous rather than discrete and
have a Gaussian prior.

At variance with standard NN models, the pattern entries
in $H_{NN}(\bsigma|\bxi)$
%(\ref{eq:marginal})
are diluted since the corresponding bipartite SG is sparse.
While standard NN retrieve
patterns sequentially (one at time), associative networks with diluted
patterns, as resulting from the marginalization 
of a diluted bipartite SG,
are able to accomplish parallel retrieval in appropriate dilution
regimes \cite{prlnoi,short,long,multilearning2}.
We note that, in contrast, diluting NNs by removing links among neurons only results in a weakening of the sequential retrieval \cite{amit}.
Beyond their applications in biology \cite{long} and potentially
large impact in artificial intelligence for their parallel processing
capabilities, associative networks with diluted patterns reveal the
inextricable link between the (sparse) bipartite network topology
and the (parallel) mode of associative network operation.
So far, diluted associative networks have been studied via replica analysis \cite{long}, for pattern-independent dilution. This setting only accounts for special structures of the underlying bipartite graph, with all degrees in each
set drawn from the same Poisson distribution. Here we
use cavity (i.e.\ belief-propagation) methods to analyze the more general
scenario where degrees in the two sets of spins have arbitrary distributions,
thus allowing for a much greater variety of bipartite network structures.

We consider an equilibrated system of $N$ binary neurons $\sigma_i = \pm 1$
at temperature (fast noise) $T=1/\beta$, with Hamiltonian
$$H(\bsigma |\bxi)=-\frac12\sum_{i,j}\sum_{\mu}^P \xi_i^{\mu}\xi_j^{\mu}\sigma_i\sigma_j,$$
where pattern entries $\{\xi_i^\mu\}$ are sparse (i.e.\ the number of
non-zero entries of a pattern is finite)
\footnote{We omit the factor $\beta$ from $H_{NN}(\bsigma|\bxi)$
%(\ref{eq:marginal}),
so that our $\beta$ corresponds to $\beta^2$ in the bipartite SG.
}%
.
 We can then use a factor graph representation of the Boltzmann weight
 as $\prod_\mu F_\mu$, with factors
\be
F_{\mu}=e^{(\beta/2)\sum_{i,j \in
    O(\mu)}\xi_i^{\mu}\xi_j^{\mu}\sigma_i\sigma_j}=\langle e^{z
  \sum_{i \in O(\mu)}\xi_i^{\mu}\sigma_i}\rangle_z,
\label{eq:z_average}
\ee
where $O(\mu)=\{i:\xi_i^{\mu}\neq 0\}$ and $z$ is a zero mean Gaussian
variable with variance $\beta$
\footnote{
Eq.~(\ref{eq:z_average}) corresponds to Gaussian $\tau_i$ in the
bipartite SG, at our redefined $\beta$; for
discrete $\tau_i$ one would average over $z=\pm \beta$ with
probability $1/2$ each.}%
.
We denote by $e_{\mu}=|O(\mu)|$ the degree of a pattern $\mu$ and by $d_i=|N(i)|$ the degree of a neuron $i$, with $N(i)=\{ \mu : \xi_i^{\mu} \neq 0 \}$. We consider random graph
ensembles with given degree distributions $P(d)$ and $P(e)$, and nonzero $\xi$'s
independently and identically distributed (i.i.d.). Conservation of links demands $N\langle d \rangle = P
\langle e \rangle$ where averages are taken over $P(d)$ and $P(e)$.
The {\em message} from factor $\mu$ to node $i$ is the cavity distribution $P_{\mu}(\sigma_j)$ of $\sigma_j$ when this is coupled to factor $\mu$ only, which we can parametrize by an effective field $\psi_{\mu \to j}$. The message from node $j$ to factor $\mu$ is the cavity distribution $P_{\setminus \mu}(\sigma_j)$ of $\sigma_j$ when coupled to all factors except $\mu$, which we can parametrize by the field $\phi_{j \to \mu}$.
The cavity equations are then \cite{cavity}
\begin{eqnarray}
P_{\mu}(\sigma_j) &=& \textrm{Tr}_{\{\sigma_k \}} F_{\mu}(\sigma_j, \{ \sigma_k \})\prod_{k \in O(\mu)\setminus j} P_{\setminus \mu} (\sigma_k),\\
P_{\setminus \nu}(\sigma_j) &=& \prod_{\mu \in N(j)\setminus \nu} P_{\mu}(\sigma_j),
\end{eqnarray}
and translate to equations for the effective fields:
\begin{eqnarray}\label{update}
&& \psi_{\mu \to j} = \tanh^{-1} \langle \sigma_j \rangle_{\mu}= \\ \nonumber && \tanh^{-1} \frac{\langle \sinh(z \xi_j^{\mu})\prod_{k \in M(\mu)\setminus j} \cosh(\phi_{k \to \mu}+z \xi_k^{\mu}) \rangle_z}
{\langle \cosh(z \xi_j^{\mu})\prod_{k \in M(\mu)\setminus j} \cosh(\phi_{k \to \mu}+z \xi_k^{\mu}) \rangle_z},\\
&&\phi_{j \to \nu} = \sum_{\mu \in N(j)\setminus \nu} \psi_{\mu \to j}.
\end{eqnarray}
These equations, once iterated to convergence, are exact on tree
graphs. They will also become exact on graphs sampled from our
ensemble in the thermodynamic limit of large $N$, because the sparsity of the
$\xi_i^\mu$ makes the graphs locally tree-like, with typical loop
lengths that diverge (logarithmically) with $N$.

\begin{figure}
\includegraphics[width=0.4\textwidth]{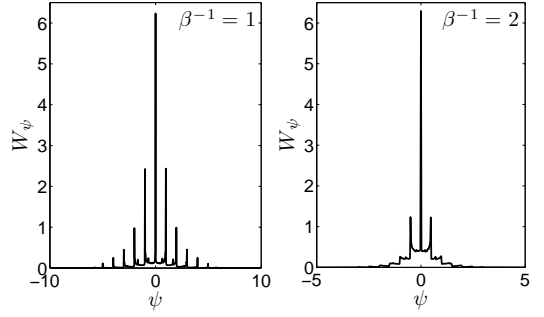}
%\hfill
%\includegraphics[width=0.23\textwidth]{istopsibreve_t.5a8c2.eps}
%\centering
%\includegraphics[width=0.23\textwidth]{istopsi_t.8a8c2.eps}
%\hfill
%\includegraphics[width=0.23\textwidth]{istopsi_t2a8c2.eps}
\caption{Histograms $W_\psi(\psi)$ of the field $\psi$ for $\alpha=8$, $c=2$ 
and $\beta^{-1}=1, 2$, as shown in figure.
%Each $\psi$ represents a $\beta\xi_i^\mu\sum_{j\in O(\mu)\notin i}\sigma_j$ Peaks at multiples of $\beta$ arise from neurons with strongly
%polarized interaction partners $\sigma_j$ where the field from factor $\mu$,  
}
\label{fig:supp}
\end{figure}

For large $N$, we can describe the solution of the cavity equations on
any fixed graph -- and hence also the quenched average over the graph ensemble and 
the nonzero pattern entries $\xi_i^\mu$ -- in terms of the
distribution of messages or fields, $W_{\psi}(\psi)$ and
$W_{\phi}(\phi)$. Denoting by $\Psi(\{ \phi_{k \to
  \mu}\},\{ \xi_k^{\mu}\},\xi_j^{\mu})$ the r.h.s.\ of (\ref{update}),
convergence of the cavity iterations then implies the self-consistency equation
\be
W_{\psi}(\psi)=\sum_{e}\frac{e P(e)}{\langle e \rangle}\langle \delta\left(
\psi - \Psi(\phi_1,...,\phi_{e-1},\xi^1,...,\xi^e) \right) \rangle
\nonumber\ee
where the average is over i.i.d.\ values of the (nonzero) $\xi^1,...,\xi^d$ and over i.i.d.\ $\phi_1,...,\phi_{e-1}$ drawn
from $W_{\phi}(\phi)$, and similarly
\be
W_{\phi}(\phi)=\sum_d \frac{d P(d)}{\langle d \rangle} \langle \delta\left( \phi - \sum_{\mu=1}^{d-1}\psi_{\mu} \right) \rangle,
\nonumber\ee
where the average is over i.i.d.\ $\psi_1,...,\psi_{d-1}$ drawn
from $W_{\psi}(\psi)$. Field distributions can then be obtained
numerically by population dynamics (PD) \cite{bethe}. For
symmetric $\xi$-distributions, a delta function at the origin for both $W_{\psi}, \ W_{\phi}$ is always a solution, and we find this to be stable at high temperatures.
At low $T$, the $\psi$ can become large (see Fig.~\ref{fig:supp}), hence also the $\phi$, and spins $\sigma_i$ will typically be strongly polarized. The fields $\beta\xi_i^\mu\sum_{j\in O(\mu)\setminus i} \xi_j^\mu\sigma_j$ then fluctuate little, and the $\psi$ as suitable averages of these fields cluster near multiples of $\beta$ (for $\xi=\pm 1$).

\begin{figure}
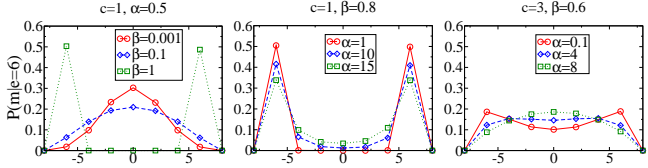

\includegraphics[width=0.162\textwidth]{sx.eps}
\includegraphics[width=0.15\textwidth]{centro.eps}
\includegraphics[width=0.15\textwidth]{dx.eps}
\caption{$P(m|e=6)$ 
above (left) and crossing (middle and right) the critical line 
for different values of $\beta$ and $\alpha$,
respectively. Full red (dashed blue and dotted green) 
curves in the middle and right panels 
refer to temperatures above (below) the critical line.
%Left: $c=1$, $\alpha=0.5$, $\beta=0.001, 0.1, 1$. Middle: 
%$c = 1$, $\alpha=1,10,15$ and $\beta = 0.8$; Right:
%$c = 3$, $\alpha=0.1,4,8$ and $\beta = 0.6$.
}
\label{fig:pm}
\end{figure}

%\subsection*{Order parameter}
Our main interest is in the retrieval properties, encoded in the 
fluctuating pattern overlaps 
$m_{\mu}= \sum_{i \in
  M(\mu)}\xi_i^{\mu}\sigma_i$. Since the joint distribution of the
$\sigma_i$ in $M(\mu)$ is $F_{\mu}(\{\sigma_i\})\prod_{i\in
  M(\mu)}P_{\setminus\mu}(\sigma_i)$, the distribution of the pattern overlap is
\be
\frac{\operatorname{Tr}_{\{\sigma_i \}}\left\langle \delta(m_{\mu}-m)\exp(\sum_{i\in M(\mu)}(\xi^{\mu}_iz+\phi_{i\to\mu})\sigma_i)\right\rangle_z}{\operatorname{Tr}_{\{\sigma_i \}}\left\langle\exp(\sum_{i\in M(\mu)}(\xi^{\mu}_iz+\phi_{i\to\mu})\sigma_i)\right\rangle_z}.
\ee
Defining this as $\mathcal{P}(m,\{\phi_{i\to\mu}\},\{\xi^{\mu}_i\})$, in the graph ensemble we have
\be
P(m)=\sum_{e}P(e)\left\langle
  \mathcal{P}(m,\phi_1,\ldots,\phi_e,\xi_1,\ldots,\xi_e)\right\rangle\;.
\ee
The average here can be read as $P(m|e)$, the overlap distribution
for patterns with fixed degree $e$. Whenever $W_\phi(\phi)=\delta(\phi)$, $P(m|e)$ is the overlap
distribution for an ``effectively isolated'' subsystem of size $e$:
the neurons storing each pattern $\bm{\xi}^\mu$ can retrieve
this independently of other patterns, even though the number of
patterns is extensive. Retrieval within each group of neurons is
strongest at low temperatures (see Fig.~\ref{fig:pm} left) as expected on
general grounds. Once nonzero $\phi$ appear neuron groups are no
longer independent: intuitively, cross-talk interference between patterns emerges.

{\em Bifurcation --} When the ``parallel processor'' solution with
zero cavity fields $\phi$ becomes unstable, a bifurcation to a different stable solutions occurs. Depending on the external parameters, this can be seen in the first or second moment of the field distribution. Expanding for small fields we get
\[
%\nonumber \small
%&&\hspace*{-2mm}
\Psi(\{ \phi_{k \to \mu}\}, \{ \xi_k^{\mu}\},
\xi_j^{\mu}) \approx 
%\\ \nonumber \small
%&&\hspace*{-4mm}
 \sum_{k \in O(\mu) \setminus j} 
%\!\!\!\!
\phi_{k \to \mu}\,
\Xi(\xi_k^\mu,\xi_j^\mu,\{\xi_l^\mu\}) 
\]
with coefficients $\Xi(\xi_k^\mu,\xi_j^\mu,\{\xi_l^\mu\})$ given by
\[
\frac{\langle \sinh(z \xi_j^{\mu})\sinh(z \xi_k^{\mu})
\!\prod_{l \in O(\mu)\setminus \{ j, k \}}\cosh(z \xi_l^{\mu})\rangle_z}
{\langle \prod_{l \in O(\mu)}\cosh(z \xi_l^{\mu})\rangle_z}\ .
\]
The self-consistency relations for the field distributions $W_{\psi}$ and
$W_{\phi}$ then show that as long as the mean fields are small, they are
related to leading order by 
\begin{eqnarray}
\langle \psi \rangle &=& \langle \phi
\rangle \sum_e P(e) \frac{e(e-1)}{\langle e\rangle}
\langle \Xi(\xi_1,\ldots,\xi_e)\rangle
\label{eq:psi_mean}
\\
\langle \phi \rangle &=&\branch_d \langle \psi \rangle
\label{eq:phi_mean}
\end{eqnarray}
%where
%\[
%\Xi = \frac{\langle \sinh(z \xi_1) \sinh(z \xi_2) \prod_{l=3}^e \cosh(z \xi_l)\rangle_z}
%{\langle \prod_{l \in M(\mu)} \cosh(z \xi_l^{\mu}) \rangle_z}
%\]
% and 
where $\branch_d = \sum_d P(d) d(d-1)/\langle d \rangle$ is one of the two branching ratios of our locally tree-like graphs, the other being $\branch_e = \sum_e P(e) e(e-1)/\langle e
\rangle$. 
If the means are zero then the onset
of nonzero fields is detected by the variances, which are related to leading order by
\begin{eqnarray}
\langle \psi^2 \rangle &=& \langle \phi^2 \rangle
\sum_e P(e) \frac{e(e-1)}{\langle e\rangle}
\langle \Xi^2(\xi_1,\ldots,\xi_e)\rangle
%
%\left(\frac{\langle \sinh(z \xi_1) \sinh(z \xi_2) \prod_{l=3}^e \cosh(z \xi_l)\rangle_z}
%{\langle \prod_{l=1}^e \cosh(z \xi_l) \rangle_z}\right)^2 
\label{eq:psi_var}\\
\langle \phi^2 \rangle 
&=& \branch_d \langle \psi^2 \rangle
\label{eq:var}
\end{eqnarray}

\begin{figure}%[h!]
\includegraphics[width=0.4\textwidth]{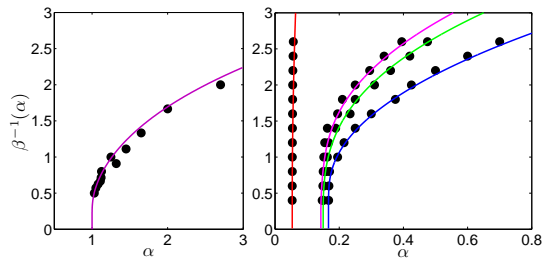}
\caption{Transition lines (theory, with symbols from PD numerics) for different
pattern degree distributions.
Left: $e\!\sim\!\mbox{Poisson}(c\!\!=\!\!1)$.
Right: changing $P(e)$ at constant $\langle e\rangle =3$; $P(e)\!=\!\delta_{e,3}$ (blue);
$P(e)=(\delta_{e,2}+\delta_{e,3}+\delta_{e,4})/3$ (green);
$P(e)=(\delta_{e,2}+\delta_{e,4})/2$ (pink); $P(e)$ power law as in
preferential attachment graphs, with $\langle e^2\rangle=21.66$ (orange).}
\label{fig:fixgrad}
\end{figure}

{\em Symmetric pattern distributions --}
When the $\xi$ are symmetrically distributed, then also the field
distributions are always symmetric and there can be no instability from growing means; cf.\ (\ref{eq:psi_mean}).
The bifurcation has to result from the growth of the variances, which from (\ref{eq:var}) occurs at $A=1$ with
\be\label{A}
A\!=\!\branch_d \sum_e P(e) \frac{e(e-1)}{\langle e\rangle}
\langle \Xi^2(\xi_1,\ldots,\xi_e)\rangle
%\branch_e
%\langle
%\left(\frac{\langle \sinh(z \xi_1) \sinh(z \xi_2) \prod_{l=3}^e \cosh(z \xi_l)\rangle_z}
%{\langle \prod_{l=1}^e \cosh(z \xi_l) \rangle_z}\right)^{\!\!2}\!\rangle
\ee
   This factorizes as $A=\branch_d A_e(\beta)$ with
%$A_d = \langle d(d-1) \rangle/\langle d \rangle$, and
the dependence on the
noise and the distribution of the $e$'s contained in the second factor $A_e(\beta)$. For $\beta \to 0$ the variance of $z$ goes to zero and $A_e(0)= 0$. For $\beta \to \infty$, the $z$-averages are dominated by large values of $z$ where $\sinh^2(z) \approx \cosh^2(z)$, so $A_e(\infty)= \branch_e$.
Hence there is no bifurcation when $\branch_d \branch_e < 1$, in agreement with the general bipartite tree percolation condition
\cite{strogatz}. For the case $P(\xi_i^\mu=\pm 1)=c/(2N)$ considered in \cite{long}, the distributions of
pattern degrees $e$ and neuron degrees $d$ are Poisson$(c)$ and Poisson$(\alpha c)$, respectively,
so $\branch_d=\alpha c$, $\branch_e=c$
and there is no bifurcation for
$\alpha c^2 < 1$.
%%which represents
%Generally, below the percolation threshold
%%of the bipartite network, in agreement with \cite{long}:
%the factor graph has no giant connected component, 
The network acts as a parallel processor here for {\em any} $T$ because the
bipartite network consists of
finite clusters of interacting spins in which there is no
interference between different patterns \cite{long}.
At higher connectivity, the critical line defined by $A=1$ indicates
the temperature above which this lack of interference persists even though
the network now has a giant connected component.
Fig.~\ref{fig:fixgrad} (left) compares theory to PD results,
where we locate the transition as the onset of nonzero second moments of the field distributions. The impact of the
transition on the overlap probability distribution of a
pattern with fixed $e$ can be seen from the PD results in 
Fig.~\ref{fig:pm} (middle and right panels). 
Crossing the transition line, parallel retrieval 
is accomplished at low temperatures, but it degrades when $\alpha$ is 
increased (see shrinking peaks in the middle panel), or $c$ 
is increased, eventually fading away for 
sufficiently large $\alpha$ and $c$ (right panel).  

One advantage of our present method is that we can easily investigate the
parallel processing capabilities of a bipartite graph with arbitrary
degrees $\{e_\mu\}$. Here we have a pattern-dependent dilution of the links 
$P(\bxi) \propto \prod_{i,\mu} P(\xi_i^\mu)\prod_\mu \delta_{e_\mu,\sum_i \mid\xi_i^\mu\mid}$
with
\be\label{costraint}
P(\xi_i^\mu)=\frac{e_\mu}{2N}(\delta_{\xi_i^\mu,1}+\delta_{\xi_i^\mu,-1})
+(1-\frac{e_\mu}{N})\delta_{\xi_i^\mu,0}
\ee
leading to $P(d)=\hbox{Poisson}(\alpha \langle e \rangle$) while
$P(e)=P^{-1}\sum_\mu \delta_{e,e_\mu}$. If we keep the mean degree fixed
$\langle e \rangle=c$, the critical point for $\beta \to \infty$ is found at
\[
\branch_d \branch_e=\alpha c (\meanv{e^2}/{c}-1)=\alpha \left[c (c-1)+\operatorname{Var}(e)\right]=1
\]
while for large $\alpha$ one obtains for the critical line
$\beta^{-1}_c(\alpha)\approx \sqrt{\alpha}\sqrt{c(c-1)+\operatorname{Var}(e)}$.
Similar results are obtained
with soft constraints $e_\mu$ on the degrees, i.e.\ by dropping the
delta function constraint in  $P(\bxi)$
%=\prod_{i\mu}P(\xi_i^\mu)$,
before $(\ref{costraint})$: one now finds
$\branch_d \branch_e=\alpha (c^2+{\rm Var}(e))$ and
$\beta^{-1}_c(\alpha)\approx \sqrt{\alpha}\sqrt{c^2+{\rm Var}(e)}$.
In both cases, the region where parallel retrieval is obtained is larger for
degree distributions with smaller variance; 
the optimal situation occurs when
all patterns have exactly the same number $c$
of non zero entries (Fig.~\ref{fig:fixgrad}, right).

{\em Non-symmetric pattern distributions} --
\begin{figure}[h!]
\includegraphics[width=0.4\textwidth]{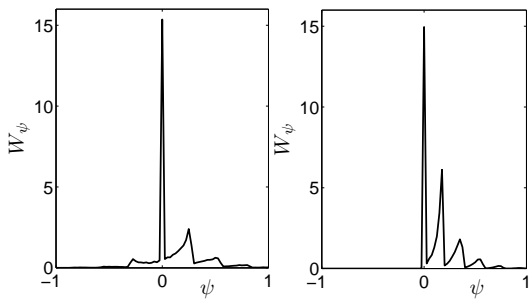}
\caption{Histogram of the fields $\psi$ in the ferromagnetic region, for
$c=1$, $\beta=1$
and different levels of bias: $a=0.9$ with $\alpha=9$ (left) and 
$a=1$ with $\alpha=8$ (right). Field distributions are obtained by PD starting from positive fields, to break the gauge symmetry. For $a=1$ (right) there are only positive fields as expected: when all patterns have positive entries there are no conflicting signals, even above the percolation threshold.
%When $a<1$ conflicting signals between different
%patterns arise above the percolation threshold.
}
\label{fig:bisto}
\end{figure}
\begin{figure}[h!]
\includegraphics[width=0.4\textwidth]{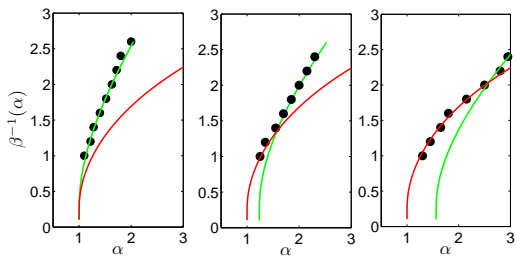}
\caption{Transition lines to growing field means (theory, green) and variances (theory, red), showing a good match to numerical PD data (dots); here $c=1$ and pattern bias $a=1, 0.95, 0.9$ from left to right. The first line to be crossed from high $T=\beta^{-1}$ gives
%Note that the physical transition line is a combination of the two curves concerning mean (green) and variance (red).
%In particular, starting from
%high temperature, the first line to be crossed gives
the physical transition.}
\label{fig:crline}
\end{figure}
To introduce a degree of asymmetry $a \in [-1,+1]$ in the pattern
distribution, we next take for the nonzero pattern entries
$P(\xi_i^\mu=\pm 1)=(1\pm a)/2$.
%
%(here we focus for simplicity on Poissonian distributions), one would take
%$$P(\xi_i^{\mu})= \frac{c}{N}(\frac{1+a}{2}\delta_{\xi_i^{\mu},+1}+\frac{1-a}{2}\delta_{\xi_i^{\mu},-1}) + (1-\frac{c}{2N})\delta_{\xi_i^{\mu},0},$$
%the updates for the mean fields become
%\begin{eqnarray}\small \nonumber
%\langle \psi \rangle &=&\E_e \langle \phi \rangle  \langle
%\frac{\langle \sinh(z \xi_1) \sinh(z \xi_2) \prod_{l=3}^e \cosh(z \xi_l)\rangle}
%{\langle \prod_{l \in M(\mu)} \cosh(z \xi_l^{\mu}) \rangle}\rangle,
%\\ \label{asimm}
%\langle \phi \rangle &=& \E_d \langle \psi \rangle
%\end{eqnarray}
%and
Evaluating  the $\xi$-average $\langle \Xi(\ldots)\rangle$ in
(\ref{eq:psi_mean}), the condition for a transition to nonzero field
means then becomes 
\be\label{bif1}
1 = a^2 \branch_d \sum_e P(e) \frac{e(e-1)}{\langle e\rangle}
\frac{\meanv{\sinh^2(z)\cosh^{e-2}(z)}_z}{\meanv{ \cosh^e(z)}_z}
\ee
At zero temperature the bifurcation occurs when $\branch_d
\branch_e = 
%\alpha c^2 = 
a^{-2}$;
when $a$ tends to zero the transition point goes to infinity and
we retrieve the symmetric case. Beyond the bifurcation, non-centered field probability
distributions (see Fig.~\ref{fig:bisto})
produce a non-zero global magnetization typical of ferromagnetic
systems. One has to bear in mind, however, that even
with a
bias in the pattern entry distribution a bifurcation to growing field variances at zero means can occur; the physical bifurcation is the one occurring first on lowering $T$. Numerical evaluation shows that both bifurcation temperatures 
increase with $\alpha$.
For large $\alpha$ one can then resort to a low-$\beta$ expansion:  $\meanv{\sinh^2(z)\cosh^{e-2}(z)}\approx\meanv{z^2}=\beta$, $\meanv{\cosh^e(z)}\approx 1$. This gives for the growing mean bifurcation condition $1\approx \branch_d \branch_e \beta a^2$ 
while for the growing variance bifurcation $1\approx \branch_d\branch_e \beta^2$.
For Poisson graphs $\branch_d\branch_e=\alpha c^2$, giving the transition lines
$\beta^{-1}_{c,1}(\alpha)\approx c^2 a^2 \alpha$ and  $\beta^{-1}_{c,2}(\alpha)\approx c
\sqrt{\alpha}$ for large $\alpha$. In the presence of a nonzero pattern bias $a$ these cross at $\alpha = 1/(ca^2)$, with the bifurcation to growing means occurring first for larger $\alpha$.
The existence of this crossing is confirmed by numerical evaluation of $(\ref{A})$ and
$(\ref{bif1})$ for finite $\alpha$ in Fig.\ $\ref{fig:crline}$.
%We remark that only
%the first line that is crossed starting from high temperature
%has physical meaning, as the hypotheses for the expansion giving the second
%line are no longer valid when the first line is crossed:
%if variances start to grow, terms depending on
%$\meanv{\psi^2}$ in (\ref{eq:var}) can no longer be neglected,
%whereas, when means grow,
%terms proportional to $\meanv{\psi}^2$ neglected
%in (\ref{eq:psi_mean}) must be accounted for.
\newline
In conclusion, we have developed a cavity/belief-propagation framework to analyse finitely connected bipartite spin glasses,
with arbitrary structure and an arbitrary degree of asymmetry in the link distribution, as well as
thermodynamically equivalent associative networks with diluted patterns.
Extensive multitasking features appear quite naturally in these systems. Our framework has enabled us to investigate their robustness for
arbitrary pattern degree distributions and asymmetry, by locating the transition surface that separates the region in $(\alpha,\beta,c)$-space
where the network is capable of parallel extensive retrieval, from the region
where pattern interference affects the network 
performance as a parallel processor.
Our results show that homogeneous degree distributions in the bipartite 
network favour parallel retrieval. 
In addition, we find that a biased distribution of the 
sparse pattern entries 
can yield a macroscopic net magnetization and shrinks
the region of parameter space where no pattern cross-talk occurs. 
However, we note that in the ferromagnetic region, pattern cross-talk 
may result in a constructive interference between patterns, which  
does not disrupt the parallel retrieval performed by the 
network.
%My interpretation would be that pattern bias adversely affects parallel 
%retrieval. Or are you thinking of a state with nonzero magnetization as 
%still performing parallel retrieval? @@
Our analysis makes contact with previous replica calculations \cite{short,long} for homogeneous graphs with symmetrically distributed links. In addition,
the cavity framework allows for straightforward extensions to general graph topology and link distributions, and may lead to a
broad range of applications, from biological to artificial systems.
%\section*{Acknowledgements}

%\vspace{0.15cm}

AB and DT acknowledge the FIRB grant RBFR08EKEV, Sapienza University 
and GNFM-INdAM for financial support.  
PS acknowledges funding from the EU under REA grant agreement nr.\ 290038 (NETADIS). Elena Agliari is acknowledged for helpful interactions.

\end{document}